\documentclass[a4paper,11pt]{article}
\usepackage{epsfig, graphicx, subfloat, float}
\usepackage{latexsym}
\usepackage{epstopdf}
\author{H. Mohseni Sadjadi\footnote{mohsenisad@ut.ac.ir} and V. Anari\footnote{v.anari@ut.ac.ir}
\\ {\small Department of Physics, University of Tehran,}
\\ {\small P. O. B. 14395-547, Tehran 14399-55961, Iran}}
\title {Early dark energy and the screening mechanism}
\begin{document}
\maketitle

\begin{abstract}
In the early Universe, dark energy may have a non-negligible contribution. If this dark energy corresponds to an early cosmological constant, it leads to a late-time huge dark energy density which is much larger than what is expected for the current expansion of the Universe. Using a conformal coupling of the quintessence to the dark matter, and based on the screening models, we propose a process to reduce the early dark energy to its ultimate value in the late time.  Early and late-time dark energies can be unified in a single component, in this way.
\end{abstract}

\section{Introduction}
The cause of the positive acceleration of the Universe \cite{acc1,acc2} is not yet clear. A cosmological constant, a scalar field, or another type of exotic field with negative pressure, modified gravity, and so on, have been proposed in the last two decades to describe this phenomenon \cite{acc4,acc5,acc6}. Despite the fine-tuning problem, the cosmological constant ($\Lambda$) is a well-known possible responsible for this acceleration \cite{wein}. Another popular candidate for dark energy is a scalar field (quintessence) \cite{quint1,quint2,quint3,quint4,quint5,quint6,quint7,quint8,quint9,quint10,quint11,quint12,quint13}. For a very slow varying or a freezing scalar field, the scalar field mimics the role of the cosmological constant in the cosmic acceleration. The absence of such a field in the results of local tests of gravitation may be attributed to the screening aspects of these models in dense regions \cite{screen}. In the symmetron model \cite{sym1,sym2}, in dense regions, the model possesses a $Z_2$ symmetry which breaks in dilute areas. This behavior has been employed to describe a mechanism to trigger the evolution of the quintessence due to the redshift of the matter density during the cosmic evolution \cite{hybrid,onset}. However, as pointed out in \cite{sep}, the symmetry breaking by lowering the scalar field potential does not directly lead to the Universe acceleration. But by considering the symmetron in the teleparallel model of gravitation \cite{onset}, or in a hybrid model \cite{hybrid}, or by coupling the symmetron to mass varying neutrinos as pointed in \cite{vah1,vah2,sam,vah3}, the $Z_2$ symmetry breaking could be responsible for the cosmic acceleration. In the latter case when massive neutrinos become non-relativistic, the quintessence becomes tachyonic and rolls down the effective potential while climbing up its potential gives. This gives rise to the Universe acceleration in the matter-dominated era \cite{vah1}.

The conformal coupling of neutrinos has recently been employed to introduce a model with early dark energy (EDE) contribution to alleviating the Hubble tension problem \cite{trod1, trod2}. In this context, the presence of a transient non-negligible EDE activated around the matter-radiation equality era, decreases the sound horizon $r_s$. This results in a larger value for the present Hubble parameter derived as $H_0\sim \frac{\theta_{*}}{r_s}$ , where $\theta_{*}$ is the angular size on the last scattering surface determined from the first cosmic microwave background peak. In the neutrino EDE model, the dark energy becomes active after the temperature reduces below the mass of the neutrino near the matter-radiation equality. This may be the answer to the  EDE coincidence problem \cite{trod1, trod2}.  In this mass-varying neutrino model, instabilities lead to neutrino nugget formation, which if properly modeled, can be the origin of the current dark matter \cite{nug}. Other models, based on the dynamics of a scalar field EDE, have also been proposed to resolve the Hubble tension such as an early frozen scalar field  ($z>3000$), which dilutes away like radiation or faster at later times \cite{kam}, and so on \cite{sc1,sc2}.  It seems that an EDE which lowers the sound horizon by $\sim\%7$ compared to the $\Lambda CDM$, may alleviate the Hubble tension \cite{hun,ten1,ten2,ten3,ten4,ten5,ten6,ten7}

Investigation of EDE as a dynamical ingredient with a small but non-negligible portion of the early Universe dates back to before the recent topics about the Hubble tension problem (see e.g. \cite{ed1}). In \cite{ed2}, by proposing a phenomenological parametrization of the density corresponding to the different eras in the cosmic evolution, it was asserted that the proposed early fractional dark energy density should satisfy $\Omega_d^e<0.06$. This dark energy may also affect the formation of massive structures \cite{ed3}.
EDE with the equation of state parameter $w=-1$ was used in \cite{ed4} to study CMB (cosmic microwave background) photons absorption by the 21cm hyperfine transition of neutral hydrogen reported by "The Experiment to Detect the Global Epoch of Reionization Signature" (EDGES) collaboration: the accelerated expansion due to the EDE can produce an earlier decoupling of the gas temperature from the radiation temperature than that in the standard model.
In \cite{ed5}, by using a parameterized uncoupled dark energy, it was shown that in the radiation-dominated epoch the EDE fractional density is restricted to be $\Omega_d^e<0.026$ in the radiation-dominated epoch and $\Omega_d^e<0.015$ in the redshift $100<z<1000$.
Recently, in \cite{ed6}, a constraint of the fractional EDE density is reported as $\Omega_d^e=0.163^{+0.047}_{-0.04}$ at the redshift $z=3357\pm200$. In many of the aforementioned papers like \cite{ed2,ed6}, the evolution and redshift of the phenomenological EDE are considered different from the late dark energy and they contribute separately to the total density.

 In \cite{ch}, to search solution for the Hubble and large-scale structure (LSS) tensions, a conformal coupling between early dark energy and dark matter has been proposed. Indeed a coupling between early dark energy and dark matter may have implications for the matter clustering and hence the LSS tension. In \cite{ch}, a quartic quintessence potential has been considered. The Chameleon exponential coupling provides a new adjustable parameter helping to tune better the model with the observable data.

In this article, unlike \cite{ed2,ed5}, we consider a common component for both the early and late dark energies. To do so, we employ the Symmetron model in which as we have mentioned before, $Z_2$ symmetry breaking reduces the quintessence potential. Instead of considering the symmetry breaking in the matter-dominated era (as is done usually to describe the late-time acceleration), we consider it at an early epoch(e.g. matter-radiation equality era). The Symmetron is only conformally coupled to the dark matter and by the redshift of the matter density, the symmetry breaking occurs and reduces the quintessence potential promptly from an initial significant value, with a non-negligible fractional density, to its reduced values in the subsequent eras.

The scheme of the paper is as follows: In the second section, we introduce the model and explain how it works by employing a simple Higgs-like potential we show how the scalar field plays the role of a bridge between the early and late-time cosmological constants. In another example, by using a more steep potential, and by choosing appropriate parameters, we illustrate our results by tracing the evolution of fractional densities and the deceleration parameter from the epoch of matter-radiation densities equality until the late time. Finally, we conclude our results in the third section.
We use units $\hbar=c=1$.

\section{Early Dark energy reduction through symmetry breaking }
We consider the following action with $Z_2$ symmetry, describing a scalar field (quintessence) interacting with other components through conformal couplings of the metric \cite{couple1,couple2}
\begin{equation}\label{1}
S=\int d^4x \sqrt{-g} \left(\frac{M_P^2R}{2}-\frac{1}{2}\partial_\mu\phi\partial^\mu\phi-V(\phi^2)\right)+S_m(\tilde{g}^{\mu \nu},\psi_j)
\end{equation}
where for the $j-th$ component: $\tilde{g}^{\mu \nu}= A_j(\phi^2)g^{\mu \nu}$. Conformal couplings $A_j(\phi^2)$'s are even functions of the scalar field. The radiation or ultra-relativistic particles do not couple to the scalar field. $M_P$ is the reduced Planck mass. $V$ is the quintessence potential.
In this work, we assume that $ A_j\neq 1$ holds only for dark matter, and for other ingredients, we take $A=1$.
By varying the action with respect to the metric, for a spatially flat isotropic homogenous space-time, we obtain the Friedmann equations
\begin{eqnarray}\label{2}
H^2&=&\frac{1}{3M_P^2}\left(\rho_r+\rho_b+A\rho_{dm}+\frac{1}{2}\dot{\phi}^2+V\right)\nonumber \\
\dot{H}&=&-\frac{1}{2M_P^2}\left(\frac{4}{3}\rho_r+A\rho_{dm}+\rho_b+\dot{\phi}^2\right)
\end{eqnarray}
The Universe is assumed to be filled with baryonic matter $\rho_b$,  radiation $\rho_r$, the quintessence dark energy $\rho_\phi$, and pressureless dark matter $A\rho_{dm}$.  Note that we have written the dark matter density in terms of a rescaled  density $\rho_{dm}$ which is conserved in the Einstein frame \cite{sym2}
\begin{equation}\label{3}
\dot{\rho}_{dm}+3H\rho_{dm}=0
\end{equation}
In our model, as we will see, before $Z_2$ symmetry breaking we have $A=1$, and $\rho_{dm}$ becomes the same as the dark matter density.

Varying the action with respect to the scalar field gives
\begin{equation}\label{4}
\ddot{\phi}+3H\dot{\phi}+V_{,\phi}+A_{,\phi}\rho_{dm} =0
\end{equation}
So, the effective quintessence potential is given by $V^{eff.}_{,\phi}=V_{,\phi}+A_{,\phi}\rho_{dm}$. The behavior of the quintessence is controlled by this effective potential and the friction term.
The radiation and the baryonic matter satisfy the continuity equations
\begin{equation}\label{5}
\dot{\rho}_r+4H\rho_r=0,
\end{equation}
and
\begin{equation}\label{6}
\dot{\rho}_b+3H\rho_b=0,
\end{equation}
respectively.  In terms of the scale factor $a$, we have $\rho_r=\rho_{r0}a^{-4}$, $\rho_{dm}=\rho_{dm0}a^{-3}$, $\rho_b=\rho_{b0}a^{-3}$. For the present time, we have taken $a_0=1$.

We intend to construct a model in which in an early era the quintessence is frozen at $\phi=\phi_{in.}$, such that its energy density is given by the constant $V(\phi_{in})$. This can be viewed as an early cosmological constant. By the redshift of dark matter, the quintessence is then activated at a specific epoch, like the matter-radiation densities equality era. This causes the reduction of the early dark energy density.
This can be realized by choosing the parameters such that for a matter density greater than a critical value $\rho_{cr}$ (which has not to be confused with the critical density for which the spatial section of the FLRW metric becomes flat), $\phi$ stays at a stable point which we take $\phi=0$. For $\rho_{dm}<\rho_{cr}$ this point becomes unstable and the field rolls down and tracks the new vacuum.
Hence, our model has an initial trivial scalar field solution which we take $\phi=0$, such that $V^{eff}_{,\phi}(0)=0$ and $V^{eff.}_{,\phi\phi}(0)>0$ for $\rho_{dm}>\rho_{cr}$. When $\rho_{dm}<\rho_{cr}$,  $V^{eff.}{_{,\phi\phi}}(0)<0$. So, we must have $A_{,\phi \phi}(0)>0$ and $V_{,\phi \phi}(0)<0$.
As for $\rho_{dm}>\rho_{cr}$, $V_{,\phi} (0)+A_{,\phi}(0)\rho_{dm}=0$, we must have $V_{,\phi}(0)=A_{,\phi}(0)=0$. Hence, after the symmetry breaking, as $V_{,\phi \phi}(0)<0$, the potential reduces. This may decrease the early non-negligible quintessence density to the expected value at the late time.

This problem is somehow similar to the spontaneous scalarization studied in black holes: where a nonminimal coupling of a scalar field to gravity may trigger an instability leading to a no trivial scalar field solution \cite{scalar1,scalar2}.

Note that cosmic acceleration occurs when
\begin{equation}
\dot{H}+H^2>0 \to 2V-A\rho_{dm}-\rho_b-2\dot{\phi}^2>0
\end{equation}
So, the fuel of the acceleration is provided by the positive potential. Although the early huge initial potential is reduced by the spontaneous symmetry breaking, it may be still enough to result in the late time  acceleration.

To show how the model works, and how the quintessence evolves, as a preliminary example we take a Higgs-like potential:
\begin{equation}
V=-\frac{1}{2}\mu^2 \phi^2+\frac{\lambda}{4}\phi^4+\Lambda^e
\end{equation}
$\Lambda^e$ plays the role of an early cosmological constant. We employ $\Lambda$ to denote the dark energy density corresponding to the cosmological constant. The conformal factor which couples the quintessence to dark matter is given by
\begin{equation}\label{A_phi}
A=1+\frac{\phi^2}{2M^2}+\mathcal{O}\left(\frac{\phi^4}{M^4}\right)
\end{equation}
We remind that the baryonic matter is not coupled to the quintessence. The effective potential is obtained as
\begin{equation}
V^{eff.}=\frac{1}{2}\left(\frac{\rho_{dm}}{M^2}-\mu^2\right)\phi^2+\frac{\lambda}{4}\phi^4+\Lambda^e
\end{equation}
The critical density is
$\rho_{cr}=\mu ^2 M^2$. Before the symmetry breaking, the fractional dark energy density is
\begin{equation}
\Omega_\phi=\frac{\rho_\phi}{3M_P^2H^2}=\frac{\Lambda^e}{\Lambda^e+\rho_{dm}+\rho_r+\rho_b}
\end{equation}

After the symmetry breaking, the new vacuum instead of $\phi=0$ is located at
\begin{equation}
\phi^2=-\frac{1}{\lambda}\left(\frac{\rho_{dm}}{M^2}-\mu^2\right),
\end{equation}
at which the potential is given by
\begin{equation}
V=\Lambda^e-\frac{\mu^4}{4\lambda}+\frac{\rho_{dm}^2}{4\lambda M^4}
\end{equation}
By using the continuity equation, we obtain
\begin{equation}
V=\Lambda^e-\frac{\mu^4}{4\lambda}+\frac{\mu^4}{4\lambda}{\rho_{dm}^{eq.}}^2\left(\frac{a_{eq.}}{a}\right)^6
\end{equation}
Where we have assumed that the symmetry breaking occurred at $a=a_{eq.}$, the epoch of matter-radiation equality. At this time the total energy density of pressureless matter equals the radiation energy density:  $\rho_m:=\rho_{dm}+\rho_b=\rho_r$.
For $a_{eq.}\ll a$, and taking by $\Lambda^e\simeq \frac{\mu^4}{4\lambda}$ we find
\begin{equation}
V\simeq\Lambda^e-\frac{\mu^4}{4\lambda}=\Lambda\ll \Lambda^e
\end{equation}
Where $\Lambda$ is considered the same as the dark energy density in the $\Lambda CDM$ model, $\Lambda=2.4\times 10^{-11} eV^4$. As the early dark energy is not negligible we expect $\Lambda\ll \Lambda^e$.  Therefore, we have taken $\Lambda^e\simeq\frac{\mu^4}{4\lambda}$, implying that the $Z_2$ symmetry breaking by activating the scalar reduced significantly the initial cosmological constant.

\begin{figure}[H]
	\centering
	\includegraphics[height=6cm]{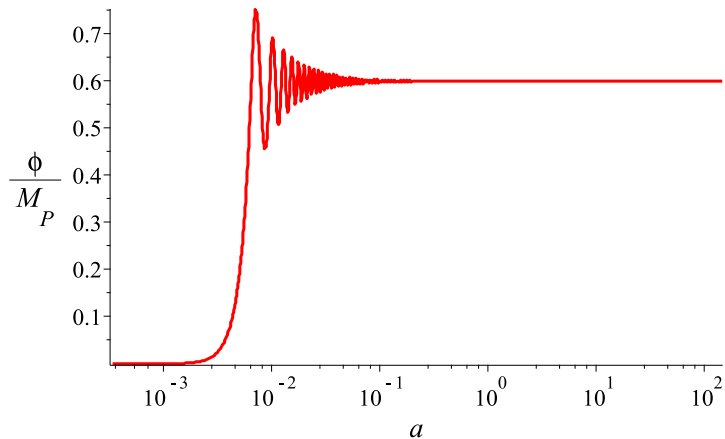}
	\caption{Quintessence evolution for $r=0.03$ and $\hat{\mu}=1$ from $\hat{\phi}=0$ at $z=3749$ to $\hat{\phi}=0.6$ in the late time }
	\label{fig_1}
\end{figure}

To illustrate this result we employ dimensionless parameters:
$\hat{H}=\frac{H}{H_{eq.}}$, $\hat{\phi}=\frac{\phi}{M_P}$, $\hat{\mu}^2=\frac{\mu^2}{H_{eq.}^2}$, $\hat{M}=\frac{M}{M_P}$, $\hat{\lambda}=\frac{M_P^2}{H_{eq.}^2}\lambda$, and $\hat{\rho}_i=\frac{\rho_{i}}{3M_P^2H_{eq.}^2}$.  In the matter-radiation equality epoch we have $\rho_r^{eq.}=\rho_{dm}^{eq.}+\rho_b^{eq.}$. In terms of the aforementioned dimensionless parameters we have:
\begin{eqnarray}
&&\hat{\rho}_{dm}'+3\hat{\rho}_{dm}=0\nonumber \\
&&\hat{\rho}_b+3\hat{\rho}_b=0\nonumber \\
&&\hat{\rho}_r'+4\hat{\rho}_r=0\nonumber \\
&&\hat{H}^2\hat{\phi}''+\hat{H}\hat{H}'\hat{\phi}'+3\hat{H}^2\hat{\phi}'-\left(\hat{\mu}^2\hat{\phi}-\hat{\lambda}\hat{\phi}^3-
3\hat{\rho}_m\frac{\hat{\phi}}{\hat{M}^2}\right)=0\nonumber \\
&&-2\hat{H}\hat{H}'=\hat{H}^2\hat{\phi}'^2+3\left(1+\frac{\hat{\phi}^2}{2\hat{M}^2}\right)\hat{\rho}_{dm}+4\hat{\rho}_r+3\hat{\rho}_b
\end{eqnarray}
where the prime denotes the derivative with respect to $\ln a$. We assume that the symmetry breaking happened during the era of equality of matter and radiation densities. In the following, we denote the contribution of dark energy to the total energy at the time of equality with $r$
\begin{equation}
r:=\Omega_\phi^{eq.}=\frac{\Lambda^e}{\Lambda^e+2\rho_m^{eq.}}
\end{equation}
In terms of $r$, we find
$\hat{\rho}^{eq.}_m=\hat{\rho}_r^{eq.}=\frac{1-r}{2}$, $\hat{\Lambda}^e=r$.
In this era, the positive acceleration happens only for $r\geq \frac{3}{7}$. If the quintessence becomes active in this era, two of the system parameters, $\lambda$, and $M$, can be estimated in terms of $\mu$ and $r$ as
\begin{equation}
\hat{M}^2\simeq\frac{5}{4}\frac{1-r}{\hat{\mu}^2},\,\,\, \hat{\lambda}=\frac{\hat{\mu}^4}{12r}
\end{equation}
By taking density parameters of matter and radiation in the present epoch as $\Omega_{m0}=0.315$ and $\Omega_{r0}=8.4 \times 10^{-5}$ and $\frac{\Omega_{b0}}{\Omega_{dm0}}=0.186$ \cite{Planck_data}, we obtain $a_{eq.}=2.67\times 10^{-4}$ (or $z=3749$). In fig.(\ref{fig_1}), the evolution of the scalar field is depicted for $r=0.03$ and $\hat{\mu}=1$. The field evolves from the initial vacuum $\hat{\phi}=0$, towards the late time  vacuum $\hat{\phi}=0.6$, which is a stable fixed attractor point at which the potential is given by $\Lambda$. A larger mass for the scalar field leads to a more prompt change in the scalar field. The scalar field tracks the minimum of the effective potential around which it rapidly oscillates with a decreasing amplitude mimicking a dark matter component.

In the above primary example, for a quintessence even with a mass of order of the Hubble parameter at the equality time $\mu\sim H_{eq.}$, the evolution of the scalar field is slow.  Therefore the initial cosmological constant is still significant after the last scattering ($z\sim 1100$) (see fig.(\ref{fig_1})).

To construct a model in which the EDE density decreases more promptly, we use another example with an even exponential potential and study the evolution of the Universe in more detail. As we will see with an appropriate choice of parameters, one can describe the late-time dark energy and EDE in the same framework. However, the model does not have an attractor solution, and we have to fine-tune the initial conditions to obtain results consistent with observations. The considered potential, consisting of an exponential term and the cosmological constant, is:
\begin{equation}\label{V}
V(\phi)=\Lambda+\Lambda^e e^{-\frac{\beta\phi^2}{2M^2}}
\end{equation}
where $\beta$ is a dimensionless constant. This is an extension of the potential employed in \cite{vah1,trod1}. Also, the conformal factor is chosen as (\ref{A_phi}). The effective potential is thus:
\begin{equation}\label{V_eff}
V^{eff.}(\phi)=\Lambda+\Lambda^e e^{-\frac{\beta\phi^2}{2M^2}}+\left(1+\frac{\phi^2}{2M^2}\right)\rho_{dm}
\end{equation}
We assume that the scalar field is initially at $\phi=0$. The squared effective mass of $\phi$ is defined as $\mu^2_{eff.}=\frac{\partial^2V^{eff.}}{\partial\phi^2}\mid_{\phi=0}$. Thus, according to (\ref{V_eff}), one finds
\begin{equation}\label{mu^2_{eff.}}
 \mu^2_{eff.}=-\frac{\beta\Lambda^e}{M^2}+\frac{\rho_{dm}}{M^2}
\end{equation}
Therefore, when $\rho_{dm}<\rho_{cr}$ the symmetry breaks, where $\rho_{cr}=\beta \Lambda^e$ is the critical density. After the symmetry breaking, the evolution of $\phi$ begins and the exponential term of the potential (\ref{V}) becomes negligible quickly. To illustrate our results, we solve numerically the following set of differential equations
\begin{eqnarray}\label{set of eqs}
\dot{a}&=&Ha \nonumber\\
\dot{\rho}_{dm}&=&-3H\rho_{dm} \nonumber\\
\dot{\rho}_r&=&-4H\rho _r \nonumber\\
\dot{\rho_b}&=&-3H\rho_b\nonumber \\
\ddot{\phi}&=&-3H\dot{\phi}-V_{,\phi}(\phi)- A_{,\phi}(\phi)\rho_{dm} \nonumber\\
H^2&=&\frac{1}{3M_P^2}( \frac{1}{2}\dot{\phi}^2+V(\phi) +A(\phi)\rho_{dm}+\rho_r+\rho_b)
\end{eqnarray}
For this purpose, we define dimensionless time $\tau$ as $\tau=tH_0$ where $H_0$ is the present Hubble parameter (the Hubble parameter at $a = 1$). The initial conditions are set at $\tau=0$ when the redshift is $z_{in.}=4.5\times 10^4$, i.e. in radiation dominated era. So, at $\tau=0$:
\begin{eqnarray}\label{IC_1}
a(0)&=&\frac{1}{z_{in.}+1} \nonumber\\
\rho_{dm}(0)&=&3\Omega_{dm0}(1+z_{in.})^3M_p^2H_0^2 \nonumber\\
\rho_{b}(0)&=&3\Omega_{b0}(1+z_{in.})^3M_p^2H_0^2 \nonumber\\
\rho_r(0)&=&3\Omega_{r0}(1+z_{in.})^4M_p^2H_0^2
\end{eqnarray}
In addition, we choose the initial conditions for the scalar field:
\begin{eqnarray}\label{IC_phi}
&&\phi(0)=0 \nonumber\\
&&\dot{\phi}(0)=10^{-5}H_0 M_P
\end{eqnarray}
and the parameters of the model as:
\begin{eqnarray}\label{parameters}
&&\Lambda =3\Omega_{\phi 0}M_P^2H_0^2 \nonumber\\
&&\Lambda^e =3rM_P^2H_{eq.}^2 \nonumber\\
&&\beta = \frac{\rho_{dm}(0)}{\Lambda^e}
\end{eqnarray}
Here, the expression for $\beta$ is obtained from $\mu^2_{eff.}=0$ in (\ref{mu^2_{eff.}}). Moreover, $\Lambda$ and $\Lambda^e$ are chosen such that $\Omega_{\phi 0}=1-\Omega_{dm0}-\Omega_{b0}-\Omega_{r0}$ and $\Omega_{\phi}^{eq.}=r$. Finally, we choose $r$ the same as the previous example, i.e. $r=0.03$, and take $M=2M_P$.\\
The initial value of  density parameters (defined as $\Omega=\frac{\rho}{3M_P^2 H^2}$) are derived from (\ref{IC_1}) and (\ref{IC_phi}) as:
\begin{equation}
\Omega_r^{in.} =0.923, \,\,\, \Omega_{dm}^{in.} =0.064,\,\,\,  \Omega_b^{in.}=0.013 \,\,\ \Omega_\phi^{in.} =8.25\times10^{-6}
\end{equation}
The initial values for $\phi$ and $\dot{\phi}$ give a negligible contribution of dark energy in the total density, but as we will it will become significant for a while near matter-radiation equality epoch. \\

\begin{figure}[H]
	\centering
	\includegraphics[height=6cm]{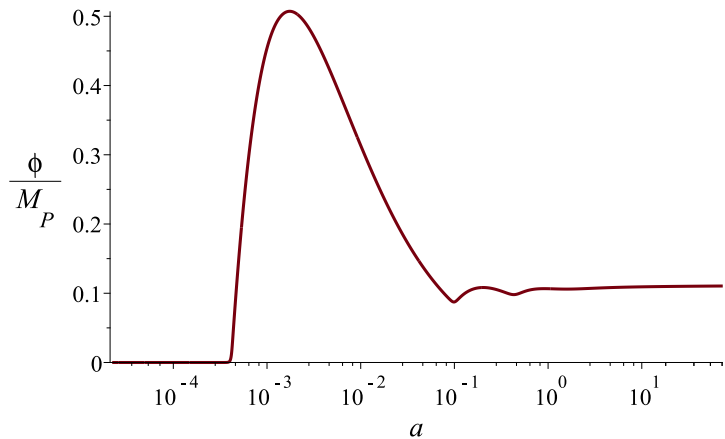}
	\caption{The scalar field in terms of $a$ , for initial conditions (\ref{IC_1}) and (\ref{IC_phi}), the parameters (\ref{parameters}), $r=0.03$ and $M=2 M_P$.}
	\label{fig_phi}
\end{figure}

In fig.(\ref{fig_phi}), $\phi$ is plotted in terms of the scale factor $a$. Initially, due to the largeness of $H$ we have a large friction term and $\phi$ evolves very slowly. At the matter-radiation equality epoch, the scalar field quickly moves down its potential until it reaches the flat region of its potential, and evolves very slowly again.

\begin{figure}[H]
	\centering
	\includegraphics[height=6cm]{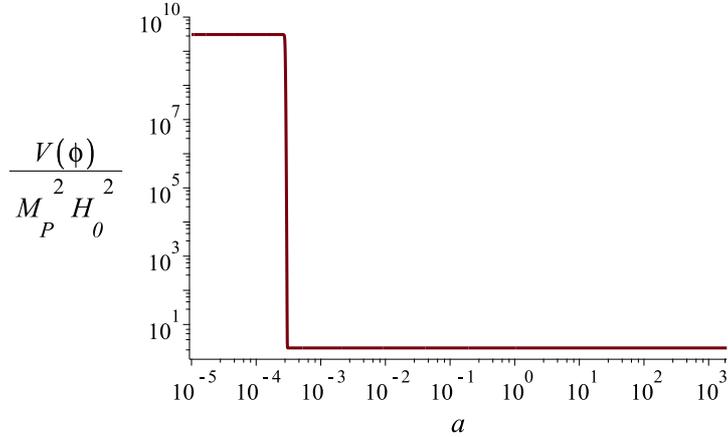}
	\caption{The potential $V(\phi)$ in terms of $a$ , for initial conditions (\ref{IC_1}) and (\ref{IC_phi}), the parameters (\ref{parameters}), $r=0.03$ and $M=2M_P$.}
	\label{fig_V}
\end{figure}

In fig.(\ref{fig_V}), the potential $V(\phi)$ is plotted in terms of $a$. Initially, $V$ behaves like a cosmological constant ($\Lambda^e$), then at the matter-radiation equality epoch, it quickly reduces to the current cosmological constant ($\Lambda$).

\begin{figure}[H]
	\centering
	\includegraphics[height=6cm]{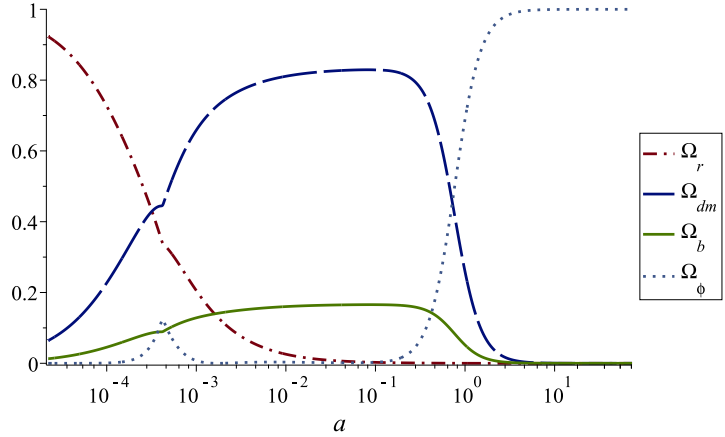}
	\caption{Relative densities in terms of $a$ , for initial conditions (\ref{IC_1}) and (\ref{IC_phi}), the parameters (\ref{parameters}), $r=0.03$ and $M=2M_P$.}
	\label{fig_Omega}
\end{figure}

In fig.(\ref{fig_Omega}), $\Omega_r$, $\Omega_{dm}$, $\Omega_b$, and $\Omega_\phi$ which are the density parameters of the radiation,
the pressureless dark matter, the baryonic matter, and the dark energy, respectively, are plotted in terms of $a$. As we can see, the universe is initially in the radiation-dominated era, where $\Omega_\phi$ is negligible.  At around the matter-radiation equality epoch ($z\simeq3749$), $\Omega_\phi$ becomes non-negligible for a short time and eventually, due to the $Z_2$ symmetry breaking, decreases before the recombination era, and then becomes again negligible. This is in agreement with other proposed models for the EDE to resolve the Hubble tension \cite{hun,ten1,ten2,ten3,ten4,ten5,ten6,ten7}. In our example, at recombination $z=1100$, we have $\frac{\rho_\phi}{3M_P^2H^2}\simeq 8\times 10^{-3}$ which is consistent with \cite{ed5}.
At the late time the fractional density increases again, and the universe transits to the dark-energy-dominated era at $z\simeq0.6$. According to the chosen parameters and initial conditions, relative densities in the present era, (corresponding to a = 1), are obtained as $\Omega_{r0}=8\times10^{-5}$, $\Omega_{dm0}=0.26$ , $\Omega_{b0}=0.05$, and $\Omega_{\phi 0}=0.69$ which are in the region expected by Planck data \cite{Planck_data}.

\begin{figure}[H]
	\centering
	\includegraphics[height=6cm]{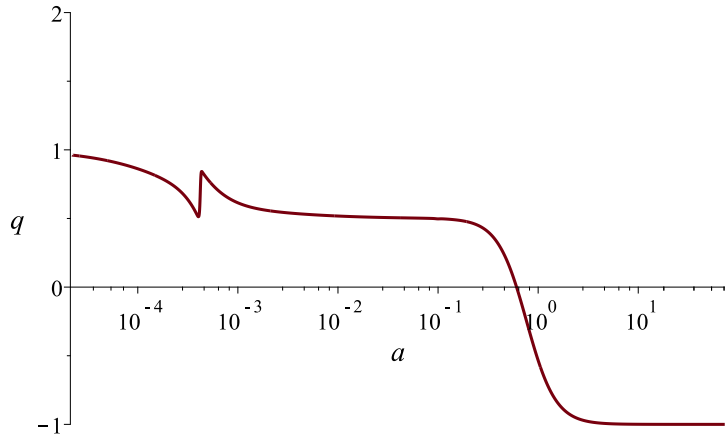}
	\caption{The deceleration parameter $q$ in terms of $a$ , for initial conditions (\ref{IC_1}) and (\ref{IC_phi}), the parameters (\ref{parameters}), $r=0.03$ and $M=2 M_P$.}
	\label{fig_q}
\end{figure}

In fig.(\ref{fig_q}), the deceleration parameter $q$ is plotted in terms of $a$. As we can see, the universe is transited from a deceleration epoch to an acceleration epoch at redshift $z=0.63$. Also, about the matter-radiation equality epoch, a different behavior than that of the $\Lambda$CDM model is observed which is due to dark energy contribution in this era.

As we add a dark energy component to the early Universe, we expect that it decreases the comoving sound horizon at the last scattering, $ r_s$, defined by
\begin{eqnarray}\label{sh}
&&r_s=\int_{z_{*}}^{\infty}\frac{c(z)}{H(z)}dz \nonumber \\
&=&\int_{z_{*}}^{\infty}\frac{c(z)}{\sqrt{\frac{1}{3M_P^2}\sum_i \rho_i}}dz \nonumber \\
&=&\frac{1}{H_0}\int_{z_{*}}^{\infty}dz\frac{c(z)}{\sqrt{\Omega_{r0}(1+z)^{4}+\Omega_{dm0}(1+z)^{3}+\Omega_{b0}(1+z)^{3}+\frac{\rho_\phi}{3M_P^2H_0^2}
}},
\end{eqnarray}
where $z_{in.}$ is the redshift of the last scattering, and $c(z)$ is the sound speed in the baryon-photon fluid. As in our model the dark energy interacts with neither baryonic matter nor radiation, we have \cite{sound}:
\begin{equation}
c_s(z)=\frac{1}{\sqrt{3}}\left( \frac{3}{4}\frac{\Omega_{b0}}{\Omega_{r0}}\frac{1}{1+z}+1\right)^{-\frac{1}{2}}
\end{equation}

A proposal to alleviate the Hubble tension, is the reduction of the sound horizon \cite{hun}. From (\ref{sh}), we find that an increase in  $H(z)$ reduces $r_s$. E.g., if the EDE density increases effectively the Hubble parameter as $H\to 1.075 H$, then $r_s$ decreases by $\% 7$ \cite{hun}(this can be accomplished if the effective role of EDE is considered as an increase of the total energy density in the integrand by the amount: $\sum \rho \to 1.156 \sum \rho$).

To get an estimation of the sound horizon one can numerically solve the set of equations (\ref{set of eqs}) equipped with the additional equation
\begin{equation}
\frac{d\psi}{dt}=\frac{1}{\sqrt{3}a}{\frac{1}{\sqrt{\frac{3\rho_b}{4\rho_r}+1}}},
\end{equation}
which is the same as  $\frac{d\psi}{dz}=-\frac{c_s(z)}{H(z)}$.  In this manner one obtains $\psi(z)$, from which the sound horizon is derived as $r_s=\psi(\infty)-\psi(z_{*})$. Repeating the same computation by ignoring EDE one obtains $r_s^{\Lambda CDM}$. By taking the last scattering redshift at $z_{*}=1100$, for the above example we obtain $\frac{r_s^{\Lambda CDM}-r_s}{r_s^{\Lambda CDM}}= 0.06$.

\section{Conclusion}
We proposed a model to unify the early non-negligible dark energy with the late-time dark energy. We employed the screening mechanism, which gives rise to a scalarization in the early Universe through $Z_2$ symmetry breaking.  This mechanism, by lowering dark energy plays like a bridge between the early and the late cosmological constants. We derived the necessary conditions for the quintessence potential and the conformal coupling.  We illustrated our results primitively by a simple quintessence with a Higgs-like potential. For this potential, the scalar field does not track the minimum of the effective potential fast enough, hence the early dark energy is still significant after the last scattering. In the second example, we employed exponential potential and studied the evolution of the Universe and its ingredients in more detail. By appropriately choosing the parameters, we have depicted the fractional energy densities and the deceleration parameter. In our numerical example, in the radiation-dominated era, the fractional dark energy density is taken as $0.03$. Near the matter-radiation equality era, the fractional dark energy density becomes significant ($\sim 12\%$), and eventually, due to the $Z_2$ symmetry breaking, diminishes before the recombination era ($\sim 0.1\%$ at $z\sim 1000$), and then is still negligible. In the matter-dominated era, the fractional density increases again and gives rise to the time acceleration at $z\sim 0.6$. In our model, the quintessence plays both the roles of the early and the late-time dark energy. In the end let us note that although our illustrative examples show how our proposed model work, the results completely depend on the chosen potential and the conformal coupling function. The parameters of the model provide us the possibility to tune the results with the observational data.

\vspace{2cm}
\textbf{Data Availability Statement}: The data generated or analyzed during this study are included in the
article.

\end{document}